# Layer-by-layer epitaxial growth of scalable WSe$_2$ on sapphire by molecular-beam epitaxy

*Masaki Nakano,*[*,†,§] *Yue Wang,*[†,§] *Yuta Kashiwabara,*[†] *Hideki Matsuoka,*[†] *and Yoshihiro Iwasa*[†,‡]

[†]Quantum-Phase Electronics Center and Department of Applied Physics, the University of Tokyo, Tokyo 113-8656, Japan.

[‡]RIKEN Center for Emergent Matter Science (CEMS), Wako 351-0198, Japan.

**ABSTRACT**: Molecular-beam epitaxy (MBE) provides a simple but powerful way to synthesize large-area high-quality thin films and heterostructures of a wide variety of materials including accomplished group III-V and II-VI semiconductors as well as newly-developing oxides and chalcogenides, leading to major discoveries in condensed-matter physics. For two-dimensional (2D) materials, however, main fabrication routes have been mechanical exfoliation and chemical-vapor deposition by making good use of weak van der Waals bonding nature between neighboring layers, and MBE growth of 2D materials, in particular on insulating substrates for transport measurements, has been limited despite its fundamental importance for future advanced research. Here we report layer-by-layer epitaxial growth of scalable transition-metal dichalocogenide (TMDC) thin films on insulating substrates by MBE, and demonstrate ambipolar transistor operation. The proposed growth protocol is broadly applicable to other TMDC, providing a key milestone toward fabrication of van der Waals heterostructures with various 2D materials for novel properties and functionalities.

After the discovery of graphene, emerging properties of two-dimensional materials with reduced thickness have attracted considerable attention both from fundamental and applied viewpoints.[1-3] Of particular interest are on monolayer group-VI transition-metal dichalcogenides (TMDC) such as $MoX_2$ and $WX_2$ ($X$ = S, Se) as a direct-gap semiconductor,[4,5] offering remarkable functionalities beneficial for electronic and optoelectronic device applications.[6,7] Moreover, broken inversion symmetry at monolayer limit with strong spin-orbit coupling gives rise to intriguing physical properties related to the valley degree of freedom,[8] providing a new platform for solid-state physics research. Those studies have been so far mainly done with mechanically-exfoliated nano-thick crystals from 'top-down' approach, while 'bottom-up' approach by thin film growth technique has been of growing significance to further exploration of physical properties and functionalities of TMDC and its heterostructures. One promising and well-developed route is to use chemical-vapor deposition (CVD),[9-11] by which high-mobility transistor operations have been already demonstrated with 4-inch wafer-scale monolayer $MoS_2$ and $WS_2$ films,[12] although large area epitaxial growth with in-plane orientation has been still a challenging issue. Another standard growth method is molecular-beam epitaxy (MBE), the state-of-the-art technique enabling growth of high-quality epitaxial thin films and heterostructures of a wide variety of materials including group III-V and II-VI semiconductors[13-15] as well as oxides[16-18] and chalcogenides,[19,20] providing groundbreaking discoveries in condensed-matter physics.

The epitaxial growth of TMDC thin films and heterostructures by MBE has a rather long history since mid '80s.[21-23] The pioneering works were done by Koma $et$ $al.$; they demonstrated epitaxial growth of $NbSe_2$ ultrathin film on a cleaved surface of $2H$-$MoS_2$ for the first time,[21] and then, succeeded in growing $MoSe_2$ epitaxial thin film on a three-dimensional $CaF_2$ (111) substrate as well.[22] In both cases, the in-plane crystallographic axes of the obtained films were highly aligned to those of the substrates despite the large lattice mismatches of about 20 %, and the authors named it 'van der Waals epitaxy', although physical properties of those films were not characterized well at that time. Recent extensive researches on TMDC have revived the concept of van der Waals epitaxy after almost thirty years, and there is indeed a growing number of papers on MBE-grown high-quality TMDC thin films being reported for the last few years.[24-28] However, those TMDC thin films have been grown on conducting graphene layers formed on SiC substrates in most cases for spectroscopic studies including angle-resolved photoemission spectroscopy or scanning tunneling microscopy, and growth of large area TMDC epitaxial thin films on insulating substrates has been less successful despite its essential importance for fundamental transport studies as well as practical electronic and optoelectronic device applications.

In this paper, we report layer-by-layer epitaxial growth of scalable $WSe_2$ thin films on insulating $Al_2O_3$ (sapphire) (001) substrates by MBE, and demonstrate ambipolar transistor operation by electrolyte gating technique. As the first demonstration, we chose $Al_2O_3$ (001) single crystal as a substrate because, (1) its hexagonal lattice might favor $c$-axis orientation of $WSe_2$, (2) it is a commercially-available insulating substrate widely used for practical electronic and optoelectronic device applications, (3) a well-defined, atomically-flat surface with regular step and terrace structure can be easily obtained by simple annealing in air, and (4) it is thermally stable up to high temperature required for growth of high-quality $WSe_2$ thin films. Figure 1a presents a top view of $WSe_2$ and $Al_2O_3$ crystals. Although the $a$-axis lattice constant of $WSe_2$ is largely different from that of $Al_2O_3$, $3 \times 3$ superlattice of $WSe_2$ and $2 \times 2$ superlattice of $Al_2O_3$ present a smaller lattice mismatch of about 4.0 %. As discussed later, it turned out that this combination enables epitaxial growth of WSe$_2$ with the relationship of WSe$_2$ (001) // Al$_2$O$_3$ (001) and WSe$_2$ [100] // Al$_2$O$_3$ [100], as illustrated in Figure 1a.

Thin film growth was carried out in an ultrahigh vacuum chamber with a base pressure below $\sim 1 \times 10^{-7}$ Pa. All the details are described in Supporting Information. An electron-beam evaporator and a standard Knudsen cell were used to evaporate W and Se, respectively. The optimized growth process is shown in Figure 1c,d. Prior to the main growth, a partly-covered amorphous buffer layer consisting of both W and Se with thickness less than monolayer was formed at room temperature, followed by the subsequent annealing at 900 $^{\circ}$C for an hour. Then, the substrate temperature was decreased down to 450 $^{\circ}$C, and the main growth was performed. When the growth was finished, the sample was annealed again at 900 $^{\circ}$C for half an hour to improve the film crystallinity, and slowly cooled down to room temperature. We note that the first process, the formation of amorphous buffer layer at room temperature and subsequent annealing at high temperature under high selenium flux prior to the main growth process was crucial for epitaxial growth of WSe$_2$ on Al$_2$O$_3$ with in-plane orientation, although the role of this amorphous buffer layer is unclear at present (for more details, see Supporting Information). Moreover, we found that this process, the formation of the amorphous buffer layer at room temperature, is broadly applicable for epitaxial growth of other TMDC thin films on Al$_2$O$_3$ substrates with high-enough crystallinity, not only for semiconducting 2H-MoSe$_2$ but also for highly-conducting 2H-NbSe$_2$ and 2H-TaSe$_2$ as well as 1T-TiSe$_2$ (see Supporting Information). Another thing to be noted is that Se should be supplied with high enough flux throughout a whole growth process to avoid having selenium vacancy in the obtained films (see Figure 1d).

An overall film growth was monitored in real time by reflection high energy electron diffraction (RHEED), which provides rich fundamental information on film growth such as an actual growth rate, time evolution of crystallinity and surface roughness of a growing film, in-plane crystallographic orientation and its relationship to that of a substrate, and so on. Figure 1e shows the time evolution of the RHEED intensity recorded during the main growth. A clear intensity oscillation was observed, indicating that $WSe_2$ thin film was grown on $Al_2O_3$ (001) substrate in the layer-by-layer mode. Each oscillation period corresponds to a growth time of monolayer $WSe_2$ as will be discussed later, and therefore by monitoring this RHEED intensity oscillation we can precisely control a layer number of a growing film. The crystallinity and the surface roughness of the film were evaluated by checking developments of the RHEED pattern. Figure 1f-i is the snapshot of the RHEED pattern along <210> azimuth of the substrate (see Figure 1a,b) taken (f) just before the main growth, (g) after the growth of one monolayer (1 ML), (h) after the growth of 10 monolayers (10 ML), and (i) after the annealing at 900 ºC. The substrate pattern changed into the film pattern with clear streaks just after the formation of 1 ML [from (f) to (g)], whose shape was kept until the end of the growth of 10 ML [from (g) to (h)], indicating two-dimensional growth without island formation throughout the growth process. After the annealing, the main streaks at $00$, $\overline{2}1$, and $2\overline{1}$ became sharper, and the dark streaks in between $00$ and $\overline{2}1/2\overline{1}$ nearly disappeared, indicating improvement of the film crystallinity and the surface roughness. Figure 1j shows the RHEED pattern along <110> azimuth of the substrate taken after the annealing, showing the different pattern from those along <210> azimuth of the substrate. This directly proves that the obtained $WSe_2$ thin film was epitaxially-grown on $Al_2O_3$ substrate with in-plane orientation. These azimuth-dependent RHEED patterns with sharp streaks did not change even when the position of the electron beam was moved over the entire film (5 mm × 5 mm in the present study) unless the substrate was rotated, proving high crystallographic homogeneity and flat surface morphology of the sample in millimeter scale. We note that we focused on the RHEED patterns to examine surface quality (crystallinity and roughness) in this paper, but we confirmed by an atomic force microscope that our $WSe_2$ thin films had very flat surfaces with a typical root-mean-squared roughness of about 0.2 nm over a few micrometer scale whether the films had grown with the buffer layer or not (see Supporting Information).

The crystallinity of the obtained 10 ML-thick $WSe_2$ epitaxial thin film was further characterized by out-of-plane and in-plane x-ray diffraction (XRD) measurements. Figure 2a displays the out-of-plane XRD pattern taken along the [001] direction of the substrate, showing the strong diffraction peak at around 13.5° and its higher-order peaks. Those peaks were assigned as the Bragg reflections from the (002) plane of the $WSe_2$ lattice, providing the *c*-axis lattice parameter of the obtained $WSe_2$ thin film to be about 1.31 nm, which is similar to the bulk value of $2H\text{-}WSe_2$ (1.30 nm). The 002 peak was accompanied by Laue oscillation, indicating high crystalline coherence along the out-of-plane direction. The thickness calculated from the Laue oscillation period was about 6.4 nm, which is consistent to the thickness of the 10 ML film as designed by the RHEED oscillation. Figure 2b shows the in-plane XRD patterns taken along the [110] (green) and [100] (blue) directions of the substrate. As expected from the azimuth dependent RHEED patterns, we confirmed clear difference in the diffraction patterns depending on the direction; the Bragg reflections from the (110) plane of the $WSe_2$ lattice were favored along the [110] direction of the substrate, whereas those from the (100) plane became dominant along the [100] direction. Figure 2c presents the reciprocal space mapping within the (*hk*0) plane, providing a whole picture of the reciprocal lattices of both $WSe_2$ and $Al_2O_3$. The 200 and 020 peaks of $WSe_2$ were nearly overlapped by the 300 and 030 peaks of $Al_2O_3$, consistent to the situation that $3 \times 3$ superlattice of $WSe_2$ nearly matches to the $2 \times 2$ superlattice of $Al_2O_3$ as pointed above. There was no apparent peak observed from different crystalline planes both in out-of-plane and in-plane XRD patterns, proving that a single phase WSe$_2$ thin film was grown in millimeter scale with the epitaxial relationship of WSe$_2$ (001) // Al$_2$O$_3$ (001) and WSe$_2$ [100] // Al$_2$O$_3$ [100]. The full-width at half-maximum (FWHM) of the $\phi$-scan at the WSe$_2$ 110 peak was about 8.8° as shown in Figure 2d, which was the smallest value that we have obtained so far.

Our MBE-grown WSe$_2$ epitaxial thin films were highly insulating, but they became highly conducting once electrons or holes were doped. To examine transport properties of our WSe$_2$ epitaxial thin films, we fabricated top-gate electric-double-layer transistors (EDLT) using ionic liquid,[29,30] N,N-diethyl-N-(2-methoxyethyl)-N-methylammonium bis-trifluoromethylsulfonyl)-imide (DEME-TFSI), and characterized transistor performances by accumulating charge carriers on a surface of WSe$_2$. All the details including device fabrication and transport measurement are described in Supporting Information. Figure 3a,b illustrates schematics of a device structure and the chemical structure of DEME-TFSI, respectively, and Figure 3c,d shows typical transfer characteristics of WSe$_2$-EDLT taken at $T$ = 220 K, demonstrating clear ambipolar operation. The enhancement of the drain current ($I_D$) was larger at negative gate voltage ($V_G$) corresponding to the hole-doped regime than positive $V_G$ corresponding to the electron-doped regime, and accordingly, the four-terminal sheet resistance ($R_s$) became lower at the hole-doped side. The minimum $R_s$ reached a few kilo-ohms, which is below the quantum resistance ($h/e^2 \sim 25.8$ k$\Omega$), implying that the sample was close to the boundary to the metallic conduction regime, although it has not been realized so far.

The number of mobile carriers and their mobilities accumulated by electrolyte gating were evaluated by Hall-effect measurements. Figure 4a-d presents the antisymmetrized transverse resistance ($R_{xy}$) as a function of external magnetic field ($\mu_0H$) at representative $V_G$ taken at $T$ =

150 K. Clear Hall signals were observed for the highly hole-doped regime ($V_G$ < -2.5 V), whereas the data were somewhat noisy at the less hole-doped regime ($V_G$ = -2.0 V) and the electron-doped regime ($V_G$ > 0 V). We note that Hall-effect measurements were successful only for in-plane oriented $WSe_2$ epitaxial thin films, and there was no discernible Hall signals detected for in-plane random $WSe_2$ thin films. The resultant sheet conductivity ($\sigma_s$) calculated from $R_s$ at zero magnetic field, the sheet carrier density ($N_s$), and the carrier mobility ($\mu$) are plotted in Figure 4e-f as a function of $V_G$ both for the hole- and electron-accumulation regimes. $\sigma_s$ was smaller at every $V_G$ than those taken at $T$ = 220 K both in the hole- and electron-doped regimes, indicating that the gate-induced carriers showed thermally-activated transport rather than band transport at the present stage as mentioned above. $N_s$ was increased linearly with increasing |$V_G$| both for the hole- and electron-doped sides, providing the capacitances of 11.7 $\mu F/cm^2$ and 5.1 $\mu F/cm^2$ for hole and electron accumulations, respectively. The maximum $\mu$ was about 3 $cm^2/Vs$ for holes and 1 $cm^2/Vs$ for electrons, respectively. The obtained 'Hall mobilities' were in the highest level among the reported 'field-effect mobilities' characterized with the first-generation CVD-grown $MoS_2$ thin films,[9-11] although approximately an order of magnitude smaller than the 'Hall mobilities' obtained for EDLT based on mechanically-exfoliated thin flakes[30] and the 'field-effect mobilities' examined with the latest-generation CVD-grown TMDC thin films.[12] We expect that further development of the growth process including choice of better substrate should lead to improvement of device performance in near future.

In summary, the present study demonstrates layer-by-layer epitaxial growth of scalable $WSe_2$ thin films on $Al_2O_3$ substrates by MBE, and verifies ambipolar transistor operation in EDLTs based on those epitaxial thin films. The obtained results are benchmarks of TMDC researches toward large area epitaxial growth of TMDC thin films for scalable devices. Moreover, the growth recipe presented in this paper is widely applicable for epitaxial growth of other TMDC thin films as well, and therefore, it should provide a fundamental route to fabrication of variety of TMDC thin films and heterostructures in a well-controlled manner, opening the door to next generation basic and applied science researches based on TMDC.

## ASSOCIATED CONTENT

### Supporting Information

The Supporting Information is available free of charge on the ACS Publications website at DOI:

Experimental details including sample fabrication and electrical transport measurements, evaluations of the buffer layer effect on crystallinity and surface roughness, and demonstration of wide applicability of the proposed growth process to epitaxial growth of other TMDC thin films (PDF)

## AUTHOR INFORMATION

### Corresponding Author

*E-mail: nakano@ap.t.u-tokyo.ac.jp

### Author Contributions

M.N., Y.W., Y.K., and H.M. grew and characterized the films. M.N. and Y.K. fabricated the devices, performed the transport measurements, and analyzed the data. M.N. and Y.I. planned and supervised the study. M.N., Y.W., and Y.I. wrote the manuscript. All the authors discussed the results and commented on the manuscript. [§]These authors contributed equally.

**Notes**

The authors declare no competing financial interests.

**ACKNOWLEDGMENT**

We are grateful to M. Kawasaki, K. S. Takahashi, Y. Kozuka, and M. Uchida for experimental help and valuable discussions for thin film growth. This work was supported by Grants-in-Aid for Scientific Research (Grant Nos. 25000003 and 15H05499) from the Japan Society for the Promotion of Science.

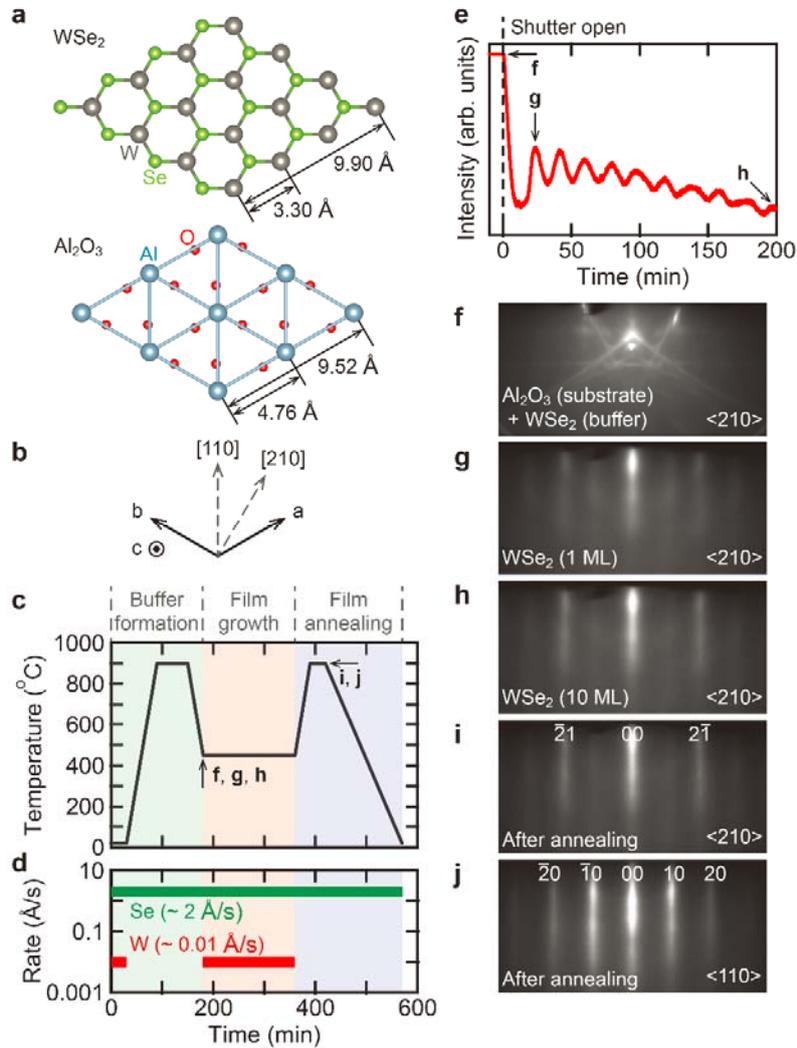

**Figure 1.** Layer-by-layer epitaxial growth of WSe$_2$ thin film on Al$_2$O$_3$ substrate. (a) Top view of WSe$_2$ and Al$_2$O$_3$ crystals with the epitaxial relationship of WSe$_2$ (001) // Al$_2$O$_3$ (001) and WSe$_2$ [100] // Al$_2$O$_3$ [100] confirmed by out-of-plane and in-plane x-ray diffraction (XRD) measurements in this study. (b) The directions of the crystallographic *a*, *b*, and *c* axes as well as the characteristic directions used for RHEED observations. (c) A schematic diagram of the growth process. (d) Typical evaporation rates of W and Se at each stage of the film growth. (e) The time evolution of the intensity of reflection high energy electron diffraction (RHEED) recorded during the film growth. (f-i) The RHEED patterns along <210> azimuth of the substrate taken at each point of the growth shown in c and e. (j) The RHEED pattern along <110> azimuth of the substrate taken after the annealing process.

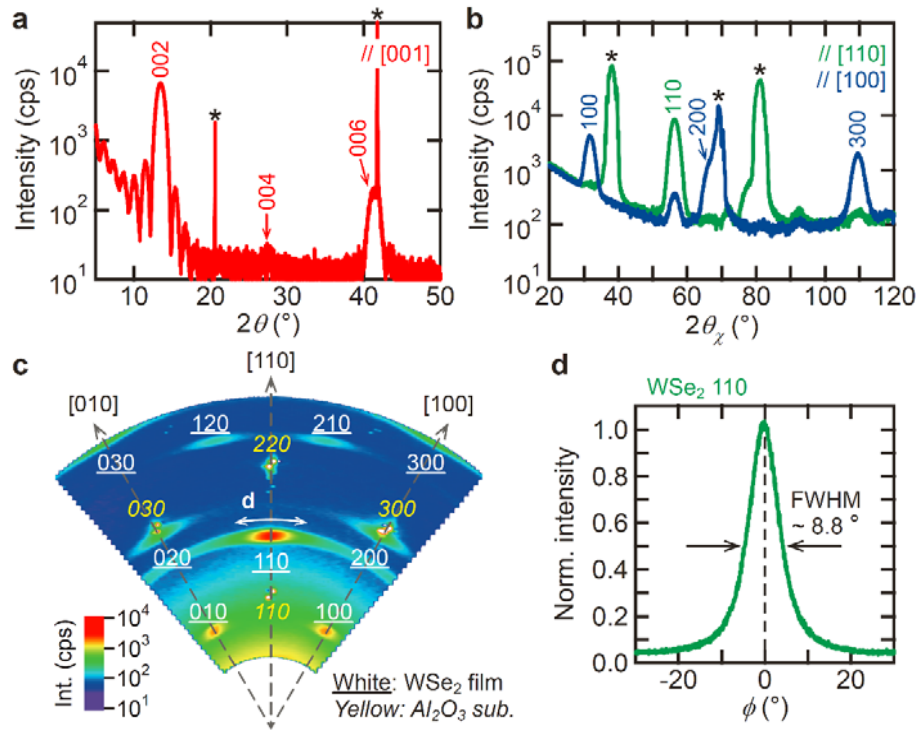

**Figure 2.** XRD patterns of 10 monolayer-thick $WSe_2$ epitaxial thin film. (a) The out-of-plane XRD pattern of the 10 monolayer-thick $WSe_2$ epitaxial thin film grown on $Al_2O_3$ (001) substrate. (b) The in-plane XRD patterns of the same sample along the [110] (green) and [100] (blue) directions of the substrate. The asterisks in a and b are the diffractions from the substrates. (c) The reciprocal space mapping of the same sample within the ($hk$0) plane. The diffraction spots labelled with white regular numbers with underlines are from $WSe_2$ thin film, and those with yellow italic numbers are from $Al_2O_3$ substrate. (d) The $\phi$ scan near the $WSe_2$ 110 peak indicated in c.

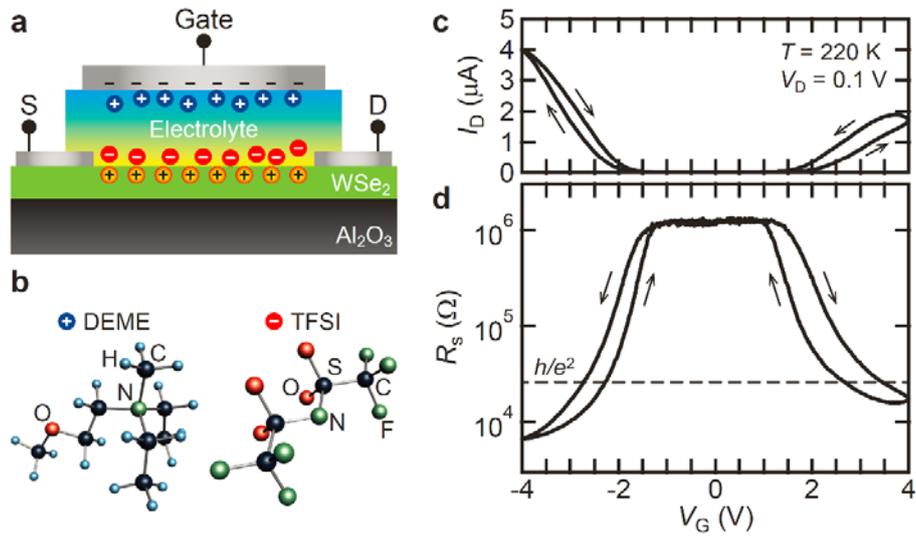

**Figure 3.** Ambipolar transistor operation in WSe$_2$ epitaxial thin film. (a) A schematic device structure of top-gate electric-double-layer transistor based on WSe$_2$ epitaxial thin film grown on Al$_2$O$_3$ substrate. (b) The chemical structure of electrolyte (ionic liquid, DEME-TFSI) used in this study. (c) The response of the drain current ($I_D$) as a function of gate voltage ($V_G$) taken at $T =$ 220 K. (d) The corresponding variation of the four-terminal sheet resistance ($R_s$) as a function of $V_G$.

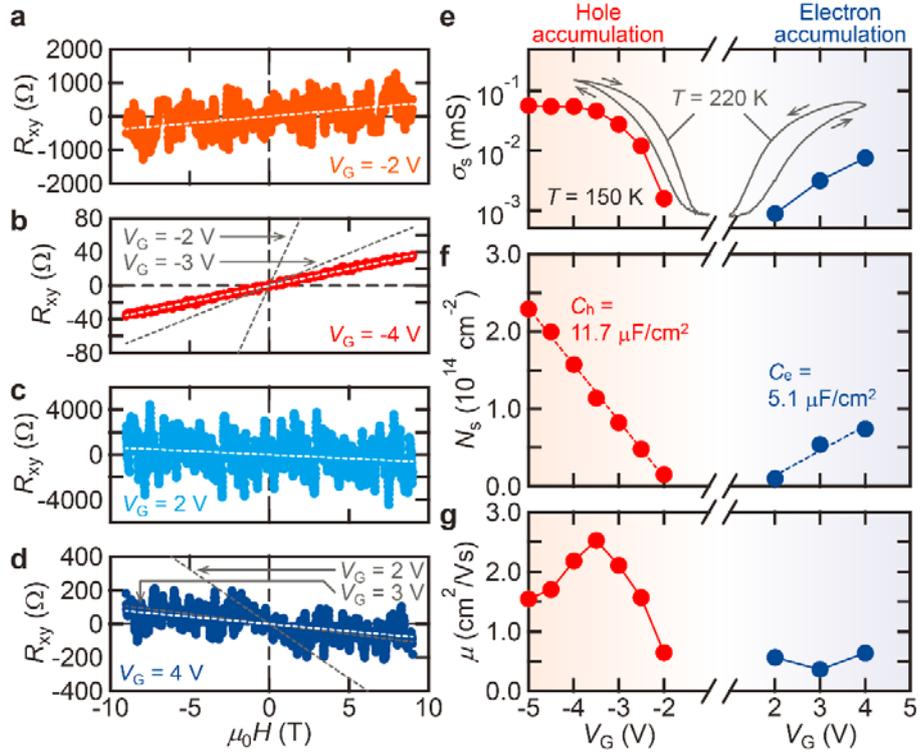

**Figure 4.** Hall-effect measurements on gated WSe$_2$ epitaxial thin film. (a-d) The antisymmetrized transverse resistance ($R_{xy}$) as a function of external magnetic field ($\mu_0 H$) at $V_G$ = -2 V (a, orange), -4 V (b, red), 2 V (c, light blue), and 4 V (d, blue) taken at $T$ = 150 K. The white broken lines are the fitted lines at each $V_G$. The gray broken lines in b and d are those at different $V_G$ for comparison. (e-g) The sheet conductivity ($\sigma_s$), the sheet carrier density ($N_s$), and the carrier mobility ($\mu$) as a function of $V_G$ at $T$ = 150 K. The gray curves in e is $\sigma_s$ at $T$ = 220 K calculated from $R_s$ shown in Figure 3d. The capacitances deduced by linear fitting of $N_s$ versus $V_G$ plots in f were 11.7 μF/cm$^2$ for hole accumulation and 5.1 μF/cm$^2$ for electron accumulation, respectively.

**TOC FIGURE:**

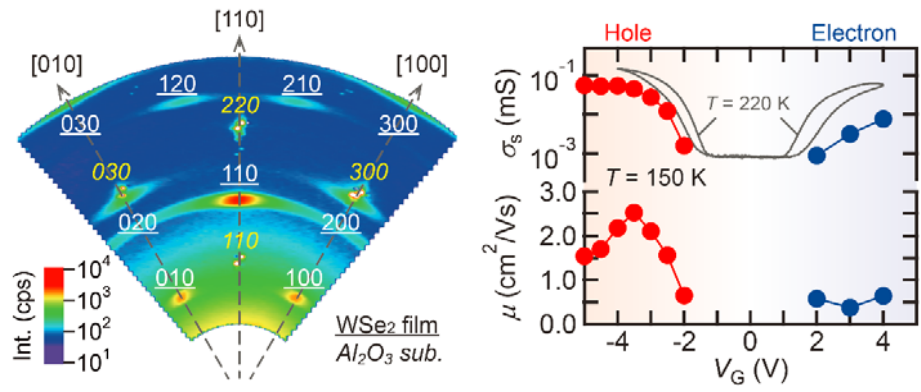